\documentclass[twocolumn,superscriptaddress,showpacs,preprintnumbers,amsmath,amssymb]{revtex4}

\usepackage{graphicx}
\usepackage{dcolumn}
\usepackage{bm}

\newcommand{\bra}[1]{{\left\langle #1 \right|}}
\newcommand{\ket}[1]{{\left| #1 \right\rangle}}

\begin{document}

\title{Faithful sharing of multipartite entanglement 
over noisy quantum channels}

\author{Soojoon Lee}\email{level@kias.re.kr}
\affiliation{
 School of Computational Sciences,
 Korea Institute for Advanced Study,
 Seoul 130-722, Korea
}
\author{Sora Choi}\email{srchoi@etri.re.kr}
\affiliation{
 Basic Research Laboratory,
 Electronics and Telecommunications Research Institute,
 Daejeon 305-350, Korea
}
\author{Dong Pyo Chi}\email{dpchi@math.snu.kr}
\affiliation{
 School of Mathematical Sciences,
 Seoul National University,
 Seoul 151-747, Korea
}
\date{\today}

\begin{abstract}
We present a protocol
in which two or more parties can 
share multipartite entanglement
over noisy quantum channels.
The protocol is based on the entanglement purification presented by
Shor and Preskill
[Phys. Rev. Lett. {\bf 85}, 441 (2000)]
and the quantum teleportation via an isotropic state.
We show that
a nearly perfect purification implies
a nearly perfect sharing of multipartite entanglement
between two parties so that
the protocol can assure a faithful sharing of multipartite entanglement
with Shor and Preskill's proof on the entanglement purification.
\end{abstract}

\maketitle

During the last two decades,
the theories on quantum communication 
protocols,
such as
quantum key distribution (QKD)~\cite{BB84,Ekert91,B92} and quantum teleportation~\cite{BBCJPW},
have considerably been developed, and have improved quantum information sciences.
Furthermore, quantum communication has almost attained to the practical stage.

A lot of quantum communication protocols~\cite{BB84,Ekert91,B92,BBCJPW,multipartite}
require perfect quantum channels,
which can conventionally be obtained from
entangled particles shared between two or more parties,
even though quantum channels are typically noisy.
Thus,
in order to succeed in a faithful quantum communication via a noisy channel,
first of all
we have to find a process to share a nearly perfect entangled state
in a given situation
by means of local quantum operations and classical communication (LOCC),
which are allowed to perform in quantum communication.
The process is called the {\em entanglement purification},
which have been studied
in several ways~\cite{BBPSSW,BDSW,VP,Horodeckis,SP,BCGST}.
In particular,
quantum error correcting codes
are closely related with
entanglement purification protocols~\cite{BDSW,SP,BCGST}.

We first review the entanglement purification protocol
presented by Shor and Preskill~\cite{SP}.
The protocol exploits the Calderbank-Shor-Steane (CSS) code~\cite{CSS},
one of the representative quantum error-correcting codes,
and has a merit that one can check the fidelity of the finally shared channel with a perfect quantum channel
before completing the protocol,
since the protocol was originally constructed
in order to prove the security of the QKD protocol proposed by Bennett and Brassard~\cite{BB84}.
Thus, if two parties successfully pass the checking procedure in the protocol, then
they can share
nearly perfect bipartite entanglements
with high probability.

We consider
the CSS code of $C_1$ over $C_2$,
which encodes $m$-qubits in $n$-qubits
and
can correct up to $t$ errors,
where $C_1$ and $C_2$ are classical linear codes such that
\begin{equation}
\{0\}\subset C_2 \subset C_1 \subset \mathbb{Z}_2^n.
\label{Eq:C_1C_2}
\end{equation}
The entanglement purification protocol
based on the CSS code
is as follows:
(1)~Alice creates $2n$ Einstein-Podolski-Rosen (EPR) pairs
    in the state $(\ket{\phi^+}\bra{\phi^+})^{\otimes 2n}$,
    where
    \begin{equation}
    \ket{\phi^+}=\frac{1}{\sqrt{2}}\left(\ket{0}\ket{0}+\ket{1}\ket{1}\right)
    \label{Eq:Bell_state}
    \end{equation}
    is one of Bell states.
(2)~Alice selects a random $2n$-bit string $b$,
    and performs a Hadamard transform on the second qubit of each EPR pair
    for which $b$ is~1.
(3)~Alice sends the second qubit of each EPR pair to Bob.
(4)~Bob receives the qubits and publicly announces this fact.
(5)~Alice selects $n$ of the $2n$ encoded EPR pairs to serve as
    check bits to test for noises.
(6)~Alice announces the bit string $b$, and which $n$ EPR pairs
    are to be check bits.
(7)~Bob performs Hadamards on the qubits where $b$ is~$1$.
(8)~Alice and Bob each measure their qubits of the $n$ check EPR
    pairs in the $\ket{0}$, $\ket{1}$ basis and share the results.
    If more than $t$ of these measurements disagree, they abort the
    protocol.
(9)~Alice and Bob make the measurements on their code qubits
    of $\sigma_z^{[r]}$ for each row $r \in H_1$ and $\sigma_x^{[r]}$ for
    each row $r \in H_2$.  Alice and Bob share the results, compute
    the syndromes for bit and phase flips, and then transform their
    state so as to obtain $m$ nearly perfect EPR pairs.

Here,
$\sigma_a^{[r]}$ is defined by
\begin{equation}
\sigma_a^{[r]}=\sigma_a^{r_1}\otimes\sigma_a^{r_2}\otimes\cdots\sigma_a^{r_n}
\label{Eq:Pauli}
\end{equation}
for a Pauli matrix $\sigma_a$, $a\in\{x,z\}$ and a binary vector $r=(r_1,r_2,\ldots,r_n)$,
and
$H_1$ and $H_2$ are parity check matrices for $C_1$ and $C_2^\perp$ respectively.
We then obtain the following lemma.

{\it Lemma 1: Shor-Preskill.}---
There exists an entanglement purification protocol
between two parties, Alice and Bob,
in which
if they have greater than an exponentially small probability of passing the test
then the fidelity of Alice and Bob's state $\rho_{AB}$ with $(\ket{\phi^+}\bra{\phi^+})^{\otimes m}$
is exponentially close to 1.

Here,
the fidelity $F$ of $\sigma$ with $\tau$ is defined by
\begin{equation}
F(\sigma,\tau)=\mathrm{tr}\left(\sqrt{\sigma^{1/2}\tau\sigma^{1/2}}\right),
\label{Eq:Def_fidelity}
\end{equation}
and we then note that
\begin{equation}
F\left(\rho_{AB},\left(\ket{\phi^+}\bra{\phi^+}\right)^{\otimes m}\right)
=\sqrt{\bra{\phi^+}^{\otimes m}\rho_{AB}\ket{\phi^+}^{\otimes m}}.
\label{Eq:pure_mixed_fidelity}
\end{equation}

In this work,
we are going to prove
the following theorem
by exploiting some appropriate LOCC and
nearly perfect bipartite entangled states
obtained from the entanglement purification protocol in Lemma~1.

{\it Theorem 1.}---
There exists a protocol
in which two parties can faithfully share a given multipartite entanglement
over noisy quantum channels.

For the detailed proof of Theorem~1,
we present some notations and two more lemmas.

Let
\begin{equation}
\ket{\Phi_d^+}=\frac{1}{\sqrt{d}}\sum_{j=0}^{d-1}\ket{j}\ket{j}
\label{Eq:d_dim_Bell_state}
\end{equation}
be one of $d$-dimensional generalized Bell states.
We remark that $\ket{\Phi_2^+}=\ket{\phi^+}$
and that when $d=2^m$
\begin{equation}
\ket{\Phi_{d}^+}_{AB}=\frac{1}{\sqrt{2^m}}\sum_{j\in \mathbb{Z}_2^m}\ket{j}_{A}\ket{j}_{B}
=\ket{\phi^+}_{AB}^{\otimes m}.
\label{Eq:2m_Bell_state}
\end{equation}

%
We now consider a one-parameter class of states in $d\otimes d$ quantum systems,
called the {\em isotropic states}~\cite{Horodeckis},
\begin{align}
\rho_F=&\frac{1-F}{d^2-1}\left(I\otimes I-\ket{\Phi_d^+}\bra{\Phi_d^+}\right)
+F\ket{\Phi_d^+}\bra{\Phi_d^+}\nonumber\\
=&\frac{d^2(1-F)}{d^2-1}\frac{I\otimes I}{d^2} +
\frac{d^2F-1}{d^2-1}\ket{\Phi_d^+}\bra{\Phi_d^+},\label{eq:isotropic}
\end{align}
with $F=\langle \Phi_d^+ |\rho_F|\Phi_d^+\rangle$.
The isotropic states $\rho_F$ have an important property
that $\rho_F$ is separable if and only if $\rho_F$ has positive partial transposition
if and only if $0\le F\le 1/d$ \cite{Horodeckis},
and furthermore several measures of entanglement for the isotropic states
can be calculated by the explicit formulas~\cite{isotropic}.
Let $\mathcal{T}_{\mathrm{iso}}$ be the $(U\otimes U^*)$-twirling operator defined by
\begin{equation}
\mathcal{T}_{\mathrm{iso}}(\rho)=\int dU (U\otimes U^*)\rho(U\otimes U^*)^\dagger,
\label{Eq:twirling}
\end{equation}
where $dU$ denotes the standard Haar measure on the group of all $d\times d$ unitary operations.
Then the operator satisfies the following two properties:
$\mathcal{T}_{\mathrm{iso}}(\rho)=\rho_{F(\rho)}$ with $F(\rho)=\bra{\Phi_d^+}\rho\ket{\Phi_d^+}$
for any state $\rho$ in a $d\otimes d$ quantum system,
and
$\mathcal{T}_{\mathrm{iso}}(\rho_F)=\rho_{F}$.
We note that $\mathcal{T}_{\mathrm{iso}}$ can be implemented
by means of LOCC~\cite{DCLB}.
Employing the isotropic states $\rho_F$ and the twirling operator $\mathcal{T}_{\mathrm{iso}}$,
we readily obtain the following lemma
which has essentially
originated from
the results in~\cite{Horodeckis}.

{\it Lemma 2.---}
Suppose that Alice and Bob share a state $\rho_{AB}$ in $d\otimes d$ quantum system,
$\mathcal{H}_A\otimes\mathcal{H}_B$,
such that
\begin{equation}\label{Eq:Phi_fidelity}
\bra{\Phi_d^+}\rho_{AB}\ket{\Phi_d^+}\ge 1-\varepsilon
\end{equation}
for some $\varepsilon >0$.
Then Alice can teleport
any pure state $\ket{\psi}$ in $\mathcal{H}_A$
to Bob in the state $\rho_{\ket{\psi}}$ satisfying
\begin{equation}\label{Eq:teleportation_fidelity}
\bra{\psi}\rho_{\ket{\psi}}\ket{\psi}\ge 1-\frac{d}{d+1}\varepsilon,
\end{equation}
by means of LOCC.

{Proof of Lemma 2. }
First, Alice and Bob transform $\rho_{AB}$ to an isotropic state $\rho_{F}$
by employing the LOCC
which can implement the $(U\otimes U^*)$-twirling operator
$\mathcal{T}_{\mathrm{iso}}$,
where $F=\bra{\Phi_d^+}\rho_{AB}\ket{\Phi_d^+}\ge 1-\varepsilon$.
Then Alice teleport a given state $\ket{\psi}$ to Bob via $\rho_{F}$,
using the standard quantum teleportation scheme.
Let $\rho_{\ket{\psi}}$ be Bob's final state.
Since the scheme produces
the fidelity 1 via a maximally entangled state $\ket{\Phi_d^+}\bra{\Phi_d^+}$ and
the fidelity $1/\sqrt{d}$ via the maximally mixed state $I\otimes I/d^2$,
it follows from Eq.~(\ref{eq:isotropic}) that
\begin{align}
\bra{\psi}\rho_{\ket{\psi}}\ket{\psi}
=& \frac{d(1-F)}{d^2-1}+\frac{d^2F-1}{d^2-1}\nonumber\\
=& \frac{Fd+1}{d+1} 
\ge 1-\frac{d}{d+1}\varepsilon.
\label{Eq:teleportation_fidelity2}
\end{align}
This completes the proof.
\hfill $\Box$

The final lemma is a generalization of Theorem 5.3 in \cite{KL} into $d$-dimensional quantum systems.

{\it Lemma 3.}---
Let $\mathcal{E}$ be a quantum operation on
a $d$-dimensional quantum system $\mathcal{H}_A$,
and $\ket{\Psi}\in \mathcal{H}_A\otimes \mathcal{H}_R$ a purification of
a state $\rho_A$ on $\mathcal{H}_A$,
where $\mathcal{H}_R$ is a reference system
such that $\mathrm{tr}_R(\ket{\Psi}\bra{\Psi})=\rho_A$.
Suppose that there is $\varepsilon > 0$ such that
\begin{equation}\label{Eq:support}
\bra{\psi}\mathcal{E}(\ket{\psi}\bra{\psi})\ket{\psi}\ge 1-\varepsilon
\end{equation}
for all $\ket{\psi}$ in the support of $\rho_A$.
Then
\begin{align}
\bra{\Psi}&\left[\left(\mathcal{E}\otimes\mathcal{I}_R\right)
\left(\ket{\Psi}\bra{\Psi}\right)\right]\ket{\Psi} \nonumber\\
&\ge 1-\left(1+d_0\cdot\max_{j\neq k}\{p_jp_k\}\right)\varepsilon,
\label{Eq:entanglement_fidelity}
\end{align}
where $d_0$ is the Schmidt number of $\ket{\Psi}$
and $\sqrt{p_j}$ are the Schmidt coefficients of $\ket{\Psi}$
with respect to the bipartite quantum system
$\mathcal{H}_A\otimes \mathcal{H}_R$.

{Proof of Lemma 3. }
By the Schmidt decomposition theorem,
$\ket{\Psi}$ can be written as
\begin{equation}\label{Eq:Schmidt}
\ket{\Psi}=\sum_{j=0}^{d-1}\sqrt{p_j}\ket{\psi_j}\otimes\ket{\phi_j}\in\mathcal{H}_A\otimes \mathcal{H}_R
\end{equation}
with $p_j\ge 0$ and
mutually orthogonal $\ket{\psi_j}$'s 
in $\mathcal{H}_A$, 
and it clearly follows that
\begin{equation}\label{Eq:rho_A}
\rho_A=\sum_{j=0}^{d-1}p_j\ket{\psi_j}\bra{\psi_j}.
\end{equation}
Then the left-hand side in Eq.~(\ref{Eq:entanglement_fidelity}) becomes
\begin{equation}\label{Eq:teleportation_fidelity2}
\sum_{j,k=0}^{d-1}\sum_{\mu}p_jp_k\bra{\psi_j}E_{\mu}\ket{\psi_j}\bra{\psi_k}E_{\mu}^{\dagger}\ket{\psi_k},
\end{equation}
where $\mathcal{E}(\sigma)=\sum_{\mu} E_{\mu}\sigma E_{\mu}^\dagger$ is the Kraus operator-sum representation of $\mathcal{E}$
with
\begin{equation}\label{Eq:normalization}
\sum_{\mu} E_{\mu}^\dagger E_{\mu} = I.
\end{equation}

For $0\le \theta \le 2\pi$,
we let
\begin{equation}\label{Eq:psi_theta}
\ket{\psi_\theta}=\sum_{j=0}^{d-1}(e^{\iota\theta})^{q_j}\sqrt{p_j}\ket{\psi_j}
\end{equation}
where $\iota=\sqrt{-1}$ and
$q_j$ are inductively defined by $q_0=0$ and $q_j=\sum_{l=0}^{j-1}q_l+1$ for $j\ge 1$,
that is, $q_j=2^{j-1}$ for $j\ge 1$.
Then
it follows from Eq.~(\ref{Eq:support}) that
\begin{align}
1-\varepsilon
\le& \bra{\psi_\theta}\mathcal{E}(\ket{\psi_\theta}\bra{\psi_\theta})\ket{\psi_\theta} \nonumber \\
=&\sum_{j,j',k,k'=0}^{d-1}\sum_{\mu}
(e^{\iota\theta})^{q_j-q_{j'}+q_k-q_{k'}}\nonumber \\
& \sqrt{p_j p_{j'} p_k p_{k'}}
\bra{\psi_{j'}}E_{\mu}\ket{\psi_k}\bra{\psi_{k'}}E_{\mu}^\dagger\ket{\psi_j},
\label{Eq:ineq01}
\end{align}
for any $0\le\theta\le 2\pi$.
Averaging uniformly the last equation in the inequality~(\ref{Eq:ineq01})
over all values of $\theta$,
from Eq.~(\ref{Eq:teleportation_fidelity2})
we obtain the following inequality:
\begin{align}
1-\varepsilon \le&
\sum_{j,k=0}^{d-1}\sum_{\mu}p_jp_k\bra{\psi_j}E_{\mu}\ket{\psi_j}\bra{\psi_k}E_{\mu}^{\dagger}\ket{\psi_k}
\nonumber\\
&+ \sum_{j\neq k}\sum_{\mu}p_jp_k\bra{\psi_j}E_{\mu}\ket{\psi_k}\bra{\psi_k}E_{\mu}^{\dagger}\ket{\psi_j}\nonumber \\
\le&
\bra{\Psi}\left[\left(\mathcal{E}\otimes\mathcal{I}_R\right)
\left(\ket{\Psi}\bra{\Psi}\right)\right]\ket{\Psi}\nonumber \\
&+ \max_{j\neq k}\{p_jp_k\}
\sum_{j\neq k}\sum_{\mu}\bra{\psi_j}E_{\mu}\ket{\psi_k}\bra{\psi_k}E_{\mu}^{\dagger}\ket{\psi_j}.
\label{Eq:ineq02}
\end{align}
We note that
\begin{equation}\label{Eq:normalization3}
\bra{\psi_k}E_{\mu}\ket{\psi_k}\bra{\psi_k}E_{\mu}^{\dagger}\ket{\psi_k}\ge 1-\varepsilon,
\end{equation}
by Eq.~(\ref{Eq:support}) in the assumption of the lemma.
Since it follows from Eq.~(\ref{Eq:normalization}) that for any $k$
\begin{equation}\label{Eq:normalization2}
\sum_{j=0}^{d-1}\sum_{\mu}\bra{\psi_j}E_{\mu}\ket{\psi_k}\bra{\psi_k}E_{\mu}^{\dagger}\ket{\psi_j}= 1,
\end{equation}
we get the following inequality:
\begin{equation}\label{Eq:ineq03}
\sum_{j\neq k}\sum_{\mu}\bra{\psi_j}E_{\mu}\ket{\psi_k}\bra{\psi_k}E_{\mu}^{\dagger}\ket{\psi_j}\le d_0\varepsilon.
\end{equation}
Hence, from the inequalities~(\ref{Eq:ineq02}) and~(\ref{Eq:ineq03})
we obtain the inequality~(\ref{Eq:entanglement_fidelity}).
Therefore, the proof is completed.
\hfill $\Box$

We remark that since $p_jp_k\le 1/4$ for all $j\neq k$
\begin{equation}\label{Eq:entanglement_fidelity2}
\bra{\Psi}\left[\left(\mathcal{E}\otimes\mathcal{I}_R\right)
\left(\ket{\Psi}\bra{\Psi}\right)\right]\ket{\Psi}
\ge 1-\frac{d_0+4}{4}\varepsilon,
\end{equation}
and that
if $\rho_A=I/d$, that is, $\ket{\Psi}$ is a pure maximally entangled state
in a $d$-dimensional quantum system
then the right-hand side in the inequality~(\ref{Eq:entanglement_fidelity})
becomes
\begin{equation}\label{Eq:entanglement_fidelity3}
1-\frac{d+1}{d}\varepsilon,
\end{equation}
and hence
with the result of Lemma~2,
we readily obtain the following corollary.

{\it Corollary 1.}---
Suppose that
Alice and Bob share a state $\rho_{AB}$ in $d\otimes d$ quantum system,
$\mathcal{H}_A\otimes \mathcal{H}_B$,
such that
\begin{equation}\label{Eq:Phi_fidelity}
\bra{\Phi_d^+}\rho_{AB}\ket{\Phi_d^+}\ge 1-\varepsilon,
\end{equation}
that
Alice prepares another state $\ket{\Phi_d^+}$,
and that
Alice teleport the second half of $\ket{\Phi_d^+}$ to Bob
via $\rho_{AB}$. Then
the state which they finally share
has the fidelity
not less than $\sqrt{1-\varepsilon}$
with $\ket{\Phi_d^+}$.

By virtue of the above lemmas,
we now prove Theorem~1.

{Proof of Theorem~1.}
For $N>m$, we let $\ket{\Psi}\in \mathcal{H}_A\otimes\mathcal{H}_{A'}$ be an $N$-qubit state
which Alice and Bob want to share
in the way that Alice and Bob possess $N-m$ and $m$ particles, respectively,
where $\mathcal{H}_A$ is an $m$-qubit system,
and let $\rho_A=\mathrm{tr}_{A'}{\ket{\Psi}\bra{\Psi}}$.

The protocol in which Alice and Bob can faithfully share $\ket{\Psi}$
is as follows:
(1)~Alice and Bob perform the entanglement purification protocol in Lemma~1,
so that they can share nearly perfect states.
(2)~Alice and Bob transform the shared state to an isotropic state by means of LOCC.
(3)~Alice prepares the state $\ket{\Psi}$,
and then they perform the standard teleportation scheme
on $m$ particles of $\ket{\Psi}$ via the isotropic state.

We now show that the above protocol can guarantee the faithful sharing of $\ket{\Psi}$.

By Lemma~1,
Alice and Bob can share $2m$-qubit state $\rho_{AB}$ such that
\begin{equation}
\bra{\Phi_{2^m}^+}\rho_{AB}\ket{\Phi_{2^m}^+}\ge 1-\varepsilon
\label{Eq:Lemma1_ineq}
\end{equation}
for some sufficiently small $\varepsilon>0$.
Thus, it follows from Lemma~2 that
Alice can teleport any $m$-qubit pure state $\ket{\psi}$
to Bob in the state $\rho_{\ket{\psi}}$ satisfying
\begin{equation}
\bra{\psi}\rho_{\ket{\psi}}\ket{\psi}\ge 1-\frac{2^m}{2^m+1}\varepsilon
\label{Eq:Lemma2_ineq}
\end{equation}
by transforming $\rho_{AB}$ to an isotropic state $\rho_F$
with $F=\bra{\Phi_{2^m}^+}\rho_{AB}\ket{\Phi_{2^m}^+}$.

Since all pure states in the support of $\rho_A$
clearly satisfy the inequality~(\ref{Eq:Lemma2_ineq}),
by Lemma~2 and Lemma~3,
we conclude that
\begin{align}
\bra{\Psi}&\left[\left(\mathcal{E}\otimes\mathcal{I}_{A'}\right)
\left(\ket{\Psi}\bra{\Psi}\right)\right]\ket{\Psi}\nonumber\\
&\ge 1-\frac{2^m}{2^m+1}\left(1+d_0\cdot\max_{j\neq k}\{p_jp_k\}\right)\varepsilon,
\label{Eq:entanglement_fidelity4}
\end{align}
where $\mathcal{E}$ is the quantum operation
representing the standard teleportation via $\rho_F$,
and
$d_0$ is the Schmidt number of $\ket{\Psi}$
and $\sqrt{p_j}$ the Schmidt coefficients of $\ket{\Psi}$
with respect to a given bipartite system $\mathcal{H}_A\otimes\mathcal{H}_{A'}$.
Therefore, since $\varepsilon$ is sufficiently small,
the proof of Theorem~1 is completed.
\hfill$\Box$

We remark that the right-hand side in the inequality~(\ref{Eq:entanglement_fidelity4})
is not less than
\begin{equation}\label{Eq:entanglement_fidelity5}
1-\frac{2^m(d_0+4)}{4(2^m+1)}\varepsilon,
\end{equation}
by the inequality~(\ref{Eq:entanglement_fidelity2}).

Since
more than two parties can share a multipartite entanglement
by sequentially executing the protocol for two parties,
we immediately obtain the following corollary.

{\it Corollary 2.}---
Several parties can faithfully share a given multipartite entanglement
over noisy quantum channels.

In conclusion,
we have presented a protocol
in which two parties can faithfully share multipartite entanglement
over noisy quantum channels, and
have shown that
a nearly perfect purification implies
a nearly perfect sharing of multipartite entanglement
between two parties.
Thus, we have finally proven that
the protocol can assure a faithful sharing of multipartite entanglement
with Shor and Preskill's proof on the entanglement purification.
For example, if Alice and Bob want to share
an $N$-qubit maximally entangled state such as
\begin{equation}
\frac{1}{\sqrt{2}}\left(\ket{0^N}+\ket{1^N}\right)
\label{Eq:example_orginal_state}
\end{equation}
so that they have $N-m$ and $m$ particles respectively,
and if the fidelity of the quantum channel obtained from Lemma~1
with a perfect quantum channel
is equal to $\sqrt{1-\varepsilon}$,
then by the proofs of the lemmas in this work
we can clearly show that
after completing the protocol
the final shared state exactly has the fidelity
\begin{equation}
\sqrt{1-\frac{3\cdot2^{m-1}}{2^m+1}\varepsilon}
\label{Eq:example_fidelity}
\end{equation}
with the original state in Eq.~(\ref{Eq:example_orginal_state}),
where the fidelity in Eq.~(\ref{Eq:example_fidelity}) is greater than $\sqrt{1-(3/2)^m\varepsilon}$.
Thus, if Alice and Bob appropriately choose the CSS code in Lemma~1
so that $\varepsilon$ is sufficiently small,
then they can share a state close to the original state.
In the similar way,
our protocol can be applied to a lot of quantum cryptographic protocols
using multipartite entanglement.
Hence, it could play a significant role in proving the security of those protocols.

S.L. acknowledges Prof. Jaewan Kim and Dr. Sangchul Oh in KIAS for useful discussions.
S.L. was supported by a KIAS Research Fund (No.~02-0140-001),
S.C. by the Korean Ministry of Planning and Budget,
and D.P.C. by
a KIAS Research Project (No.~M1-0326-08-0002-03-B51-08-002-12)
funded by the Korean Ministry of Science and Technology.

\end{document}